\documentclass[9pt,twocolumn,twoside,lineno]{pnas-new}

\templatetype{pnasresearcharticle} 

\title{Global Network Prediction from Local Node Dynamics}

\author[a]{Neave O'Clery}
\author[b]{Ye Yuan} 
\author[c]{Guy-Bart Stan}
\author[c]{Mauricio Barahona}

\affil[a]{University of Oxford}
\affil[b]{Huazhong University of Science and Technology}
\affil[c]{Imperial College London}
\correspondingauthor{\textsuperscript{a}To whom correspondence should be addressed. E-mail: oclery@maths.ox.ac.uk}
\leadauthor{O'Clery}

\keywords{Networks $|$ Dynamics $|$ Consensus $|$ Ranking $|$ localised} 

\dates{This manuscript was compiled on \today}

\begin{document}

\verticaladjustment{-2pt}

\maketitle
\thispagestyle{firststyle}

{\bf The study of dynamical systems on networks, describing complex interactive processes, provides insight into how network structure affects global behaviour. Yet many methods for network dynamics fail to cope with large or partially-known networks, a ubiquitous situation in real-world applications. 
Here we propose a localised method, applicable to a broad class of dynamical models on networks, whereby individual nodes monitor and store the evolution of their own state and use these values to approximate, via a simple computation, their own steady state solution. Hence the nodes predict their own final state without actually reaching it. 
Furthermore, the localised formulation enables nodes to compute global network metrics without knowledge of the full network structure. 
The method can be used to compute global rankings in the network from local information; to detect community detection from fast, local transient dynamics; and to identify key nodes that compute global network metrics ahead of others.
We illustrate some of the applications of the algorithm by efficiently performing web-page ranking for a large internet network and identifying the dynamic roles of inter-neurons in the \emph{C. Elegans} neural network. 
The mathematical formulation is simple, widely applicable and easily scalable to real-world datasets suggesting how local computation can provide an approach to the study of large-scale network dynamics.} 

We live in an interconnected world. From social networks to environmental sensors and autonomous robots, networked systems have become ubiquitous in our everyday lives. In order to describe such complexity, we can employ mathematical tools to model the behaviour of agents---whether they be human, organism or machine---as they communicate, process and integrate information in a networked environment. 
For instance, opinion formation evolves over time based on the belief level, with each individual updating their own opinion based that of their contacts. Eventually, as information spreads, a consensus emerges for the entire network as confidants embrace the emerging common opinion.  
Simple consensus models \cite{OlfatiSaber2007}, whereby nodes converge to a common value or opinion, fall within the general class of dynamical linear systems \cite{newman2011structure} and are employed in the study of social, biological, robotic and manufacturing systems~\cite{strogatz2001exploring, Newman2003, Egerstedt2012} with applications in distributed sensors \cite{Xiao2005}, social networks \cite{Krause1997, Sood2005} and synchronisation \cite{Barahona2003, Tanner2003, Blondel2005}. 

A key focus of recent research has been understanding and exploiting the connection between network structure and dynamics \cite{Barahona2003, Delvenne2010, oclery}. A well-known application that exploits dynamics to elucidate information about network structure is that of node ranking, which forms the basis for Google's PageRank \cite{brin1998anatomy}. Given a network of webpages connected by links, nodes iteratively update their own value by averaging the rankings of neighbouring nodes under a dynamical model. The final node ranking is derived from computation of the long-run steady state solution of this dynamical system. Other examples in the engineering and control literature focus on the ability of certain nodes to influence or decide the outcome of a dynamical system on a network \cite{Liu:2011kx}. Such analysis is critical in the field of distributed and cooperative robotics, where teams of autonomous robots or vehicles must co-ordinate their movement, and respond to external stimuli \cite{mesbahi2010graph}. Related models are also used extensively in the analysis of social networks and opinion formation~\cite{schaub2018a}. These examples can be seen as applications of generalised linear dynamics on a network, where all nodes evolve towards a final steady state that is determined by global information contained in the network~\cite{schaub2018b}. 

\begin{table}[!b]
\noindent\fbox{%
    \parbox{8.4cm}{%
    \sffamily 
   
    \center
         \begin{minipage}{8cm}
          {\bf \sffamily SIGNIFICANCE STATEMENT} \\ \\
Many technologies today rely on networks to describe complex interactions between groups of agents. Dynamical processes on networks, whereby each node evolves over time based on the state of its neighbours, have a large number of applications, including opinion models, environmental sensing and internet search. However, the analysis of large networked dynamical systems is computationally intensive, and requires potentially unavailable knowledge of the global network structure. Here we propose an efficient localised approach, whereby each node computes its own final equilibrium value using only local information collected from its own history. We show how this methodology can be used for web-page ranking, community detection, and identification of key communicator nodes in neural networks in a local manner. 
\\     \end{minipage}\par\vfill}\par
}
\end{table}

Despite much progress in this field, many existing tools and algorithms at the network-dynamics interface are ill-equipped to cope with the current explosion in data availability and network size, as well as practical, design and security challenges. For example, researchers today are regularly faced with network data containing information on millions, if not billions, of nodes. Examples include social media networks (Facebook currently has 1.5 billion active users) \cite{ellison2007benefits}, the internet \cite{strogatz2001exploring}, and biological systems such as protein interaction networks \cite{delmotte2011protein} and neural networks \cite{watts1998collective}. 
In order to compute the steady state value of the dynamics on these networks, the full dynamical system is typically simulated until its equilibrium behaviour is attained. However, due to the fact that the convergence process can be very slow~\cite{OlfatiSaber2007}, dynamical simulation is not always computationally feasible. Additionally, for many real-world practical applications, such as the internet, biological, finance and sensor networks, the full network structure is often unknown due to measurement difficulties, security concerns, and/or physical constraints. In these cases, individual agents or nodes may only have access to a limited amount of locally collected data. Hence it is often desirable, if not necessary, to employ limited local information (i.e., information on a single node, or subset of nodes) to extract \textit{both local and global} network characteristics. 

Here we propose a fully localised method, whereby individual nodes monitor the evolution of their own state as it changes, and use these values of its dynamical history to perform a simple computation that allows them to approximate their own long-term dynamics. This approach enables nodes to `predict' their own final state, and in some cases that of the whole network, well before the dynamics actually converge to an equilibrium state. 

Our method builds on work of Sundaram and Hadjicostis \cite{Sundaram:2007fk} who originally proposed a method to compute the final state using a sequence of initial state values of length equal to the rank of the observability matrix (a matrix constructed from powers of the adjacency matrix, see \cite{Liu12022013} for an overview). Yuan \emph{et al.}~\cite{Yuan2013} proposed an analogous but \textit{localised} approach by replacing the observability matrix with a local Hankel matrix 
containing only the history of state values of each node. While theoretically interesting, both approaches suffer from the fact that the observability (or, equivalently, Hankel) rank is typically large (usually equal to the network size) for most real-world networks, limiting their practical usefulness and applicability. 

Here we show that a relaxation of the localised Hankel approach generates a sequence of approximations to the steady state value for each node. Specifically, for each node and at each time step, we compute the singular value decomposition of a Hankel matrix composed of preceding state values, and use it to approximate the Hankel nullspace vector to compute an approximate steady state value for each node. Critically, this method, termed below as the `Hankel method', does not require knowledge of the full network topology, and enables nodes to not only `predict' their own long-term dynamical equilibrium or steady state, but also in many cases that of the full network.  The localised and dynamical nature of the algorithm has several desirable features.

\emph{The Hankel method is fast compared to traditional methods, and scalable to large systems.} The sequence of Hankel approximations is guaranteed to converge in a number of steps less than or equal to the Hankel rank, but in practice typically accurately approximates the steady state value in very few steps. For example, for a random (Erd\H{o}s-R\'{e}nyi) network of one thousand nodes, only three to four steps are typically needed to approximate the final value for each node. In contrast, thousands of steps may be required for convergence of the iterative dynamics to the same accuracy. We show that this approach can be used to construct a localised algorithm to compute the Google PageRank vector \cite{brin1998anatomy}, and we illustrate this for both the well-known Karate Club social network \cite{karate}, and a larger network of over $10,000$ webpages within the 2002 \emph{Stanford.edu} domain \cite{leskovec2009community}. 

\emph{The Hankel method is node-specific.} The process of predicting global information from local information is not equally possible for all nodes, i.e., some nodes can compute their own long-run equilibrium value using fewer of their own initial values. Under a consensus model \cite{OlfatiSaber2007}, whereby nodes converge to a common value or opinion, certain nodes are the first to be able to compute the full network steady state (and can communicate this information to other nodes if needed). These nodes can been seen as knowledgeable `network insiders'. We illustrate this phenomenon by considering the well-known \textit{C. Elegans} neural network, which describes the chemical and electrical wiring structure of sensory, motor and inter-neurons for this worm \cite{watts1998collective, varshney2011structural}. Assuming a simple dynamics under which neurons receive and assimilate information from their neighbours, we observe that inter-neurons tend to require fewer values of their history to compute the global outcome of the system dynamics, consistent with their role as key communicators in the network.

\emph{The Hankel method is generalisable and adaptable.} Its mathematical formulation is simple, and encompasses a large class of widely-used linear models for network dynamics. Hence, we can exploit its localised and computational efficiency for a range of applications. As illustrations, we show below that each node can not only predict its own final value or ranking, but also detect clustering of intermediate values before an equilibrium state is reached. This behaviour is relevant to the detection of modular structure in networks based on the dynamics of random walker travelling from node to node. Transient clustering of the dynamics is indicative of modular community structure, as random walkers become trapped in dynamical `basins' \cite{fortunato2010community, Lambiotte09laplaciandynamics, Delvenne2010}. Such analysis can illuminate, for example, functional groups of proteins in proteome networks, which can be associated with different diseases or biological processes \cite{jonsson2006cluster}.

\begin{figure}[t!]
\centering
\includegraphics[width=8cm]{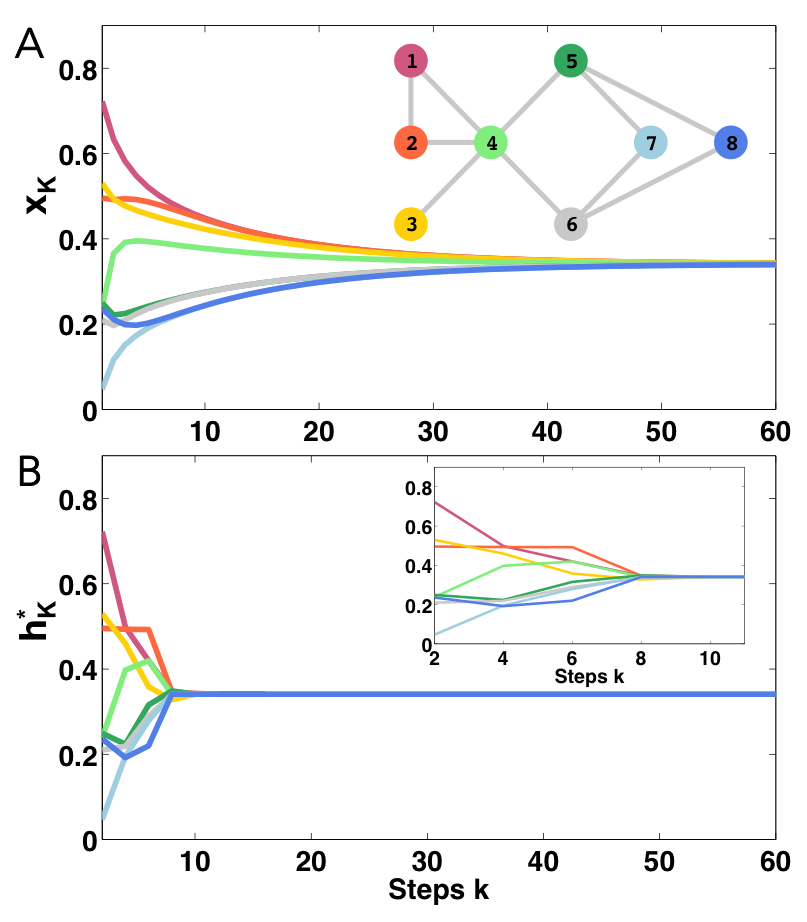}
\caption{\footnotesize \emph{Predicting the final value of Laplacian consensus dynamics.} \textbf{(A)} Laplacian consensus dynamics of the network shown in the inset for a random initial condition (time traces correspond to node colour). The iterative dynamics reaches the consensus value slowly beyond $\sim$ 60 steps. \textbf{(B)} The sequence of approximate consensus values computed via the Hankel method~\eqref{eqndecent_main} for each node converges to the true consensus value in less than 10 steps (detected with tolerance $\epsilon=10^{-4}$). The inset is a blow-up of the main figure showing that each node approximates its final value in a different number of steps; in all cases less than $2(\Delta_r+1)$ with $\Delta_r=\{6,6,5,4,5,5,4,4\}$ for nodes $r=1,...,8$, respectively~\eqref{eqnfv}.}
\label{fighank}
\end{figure}

\section*{Localised, Finite-time Computation of Network Steady States  \label{decentye}}

We will consider a generalised model of consensus dynamics whereby nodes can both store their own `value' or opinion at any point in time, and periodically learn their neighbours values in order to `update' their own opinion. 
Mathematically, if ${\bf{x}}_{k} \in \mathbb{R}^n$ is a vector containing the value or state of each node at time-step $k$, all future node values may be described by the iterative process
\begin{equation}
\label{eqnconsdis_main}
{\bf{x}}_{k+1}=W{\bf{x}}_k
\end{equation}
where ${\bf{x}}_{0} \in \mathbb{R}^n$ contains the initial condition for the dynamics, and the weight matrix $W \in \mathbb{R}^{n\times n}$ describes the inter-node relationships giving rise to the dynamical system. All nodes asymptotically (i.e. eventually) converge to the same (consensus) steady state solution under two conditions: $W{\bf 1}={\bf 1}$ (where ${\bf 1}$ corresponds to a vector of ones) and all other eigenvalues of $W$ are within the unit disk. 

Depending on the situation, $W$ can take several forms. One of  the most common forms is that of the Laplacian consensus dynamics \cite{OlfatiSaber2007}, which describes the process by which a node of a network updates its values based on that of its neighbours. In this case, $W=I-\omega L$, where $L=D-A$ is the Laplacian matrix with $A$ being the adjacency matrix of the network with entries $i,j$ containing the weight of the edge between node $i$ and node $j$, and $D$ being a matrix of zeros everywhere, except on the diagonal where it contains the corresponding node out-degree. For a directed network, if (i) the graph is balanced and strongly connected, and (ii) $0 < \omega < 1/\max(D)$, then all variables in $\bf{x}_k$ asymptotically reach a shared average consensus value. Another key example is that of random walker probability dynamics \cite{Lambiotte09laplaciandynamics}, where $W=AD^{-1}$ and the final state vector represents the long-term transition probabilities of a random walker transitioning through each node. Random walker dynamics on a network forms the basis of a number of algorithms including Google's PageRank \cite{brin1998anatomy}, and community detection algorithms \cite{fortunato2010community, Delvenne2010}.  

An example network is shown in the inset of Figure~\ref{fighank}~A, where we simulate Laplacian consensus dynamics as described above, and observe that, despite the small size of the network ($n=8$ nodes), the node dynamics do not converge to a consensus value (for a random initial condition) until step $k\approx 60$. Indeed, in all cases, given the conditions on $W$ above, the rate of asymptotic convergence for such dynamics is upper bounded \cite{OlfatiSaber2007}, yet there is no lower bound to this rate of convergence, i.e., there is no mathematical limit on the number of iterations or steps this could take.

Sundaram and Hadjicostis \cite{Sundaram:2007fk, Sundaram2007} have shown that individual nodes, under certain conditions, can compute their own final state value in a finite number of steps equal to the rank of a node-specific observability matrix (composed of matrix powers of $W$, see SI and \cite{Liu12022013} for a review). However, in most cases, the observability rank is equal to the size of the network, and nodes require knowledge of the full network structure---an assumption not likely to be met for many real-world applications. Yuan \emph{et al.} \cite{Yuan2013} proposed an analogous but localised methodology, employing a `local' Hankel matrix in place of the `global' observability matrix. Hankel matrices are square matrices in which each ascending skew-diagonal from left to right is constant, and are frequently employed in control theory. 
Specifically, if $x_i(r)$ for $i=0,1,2,\ldots$ are the state values associated with node $r$, then the Hankel matrix $H_k^{(r)} \in \mathbb{R}^{k \times k}$ of size $k$ for node $r$ is defined as
$$
H_k^{(r)} =
\left[ {\begin{array}{cccc}
x_1(r)-x_0(r) & \hdots & x_k(r)-x_{k-1}(r)\\
x_2(r)-x_1(r) & & x_{k-1}(r)-x_{k-2}(r)\\
\vdots & & \vdots \\
x_k(r)-x_{k-1}(r) &   \hdots & x_{2k-1}(r)-x_{2k-2}(r) 
 \end{array} } \right].   
$$
This matrix is increased in size (via addition of rows and columns) until its singular value decomposition \cite{golub2012matrix} includes a single zero eigenvalue and it has dropped rank at $k=\Delta_r+1$, i.e., $\text{rank}\left(H_k^{(r)}\right)=\Delta_r$. The steady state value can be then computed by node $r$ using the formula (see \cite{Yuan2013}):
\begin{equation}
\label{eqnfv}
h^*(r)=\frac{[x_0(r)\hdots x_{\Delta_r}(r)]{\bf v}_{\Delta_r+1}^{(r)}}{{\bf 1}^T{\bf v}_{\Delta_r+1}^{(r)} }.
\end{equation}
where ${\bf v}_{\Delta_r+1}^{(r)} \in \mathbb{R}^{\Delta_r+1}$ is the single nullspace vector of $H_{\Delta_r+1}^{(r)}$ and $x_0(r),\hdots, x_{\Delta_r}(r)$ are successive values of node $r$. 
A proof may be found in \cite{Yuan2013}, and is based on a Jordan decomposition \cite{golub2012matrix} of $W$ and a Vandermonde decomposition \cite{golub2012matrix} of $H_{\Delta_r+1}^{(r)}$. 
When identical symmetry exists both in the graph and initial condition, the rank of $H_{k}^{(r)}$ does not accurately reflect the number of steps to compute the final value~\cite{hend3, Yuan2013}. Yet, for a random initial condition this is highly unlikely to be an issue. 

As mentioned above for the observability method, the number of steps required for each node, given by the corresponding Hankel rank, of most common classes of complex networks is usually trivial (i.e., it equals the graph size $n$) \cite{neavephd}. This implies, firstly, that there is a prohibitive cost for very large networks with millions of nodes in computing the steady state value using the above approach, and, secondly, that all nodes compute the final value in the same number of state values---there is no node-specific predictive advantage. There is also a significant limitation to the accuracy of this approach as the graph size grows due to the fact that the entries of the Hankel matrix tend towards zero as the system converges to the consensus value, i.e., $x_k(r)-x_{k-1}(r) \to 0$ as $k \to \infty$. Determining the matrix rank accurately is therefore numerically unstable for large matrices \cite{oclery}, and particularly so for those composed of small numbers. 

We exploit the properties of the singular value decomposition to propose a relaxation of this approach. Specifically, we use the singular vector corresponding to the smallest singular value $\sigma_1$ to approximate the Hankel nullspace vector for increasing Hankel size $k$. More specifically, we compute a sequence of approximations to the steady state solution for node $r$ for steps $k\leq\Delta_r+1$ such that 
\begin{equation}
\label{eqndecent_main}
h_{k}^*(r)=\frac{[x_0(r)\hdots x_{k-1}(r)]{\bf v}_k^{(r)}}{{\bf 1}^T{\bf v}_k^{(r)} }
\end{equation}
where ${\bf v}_k^{(r)}$ is the singular vector corresponding to the smallest singular value of the Hankel matrix $H_k^{(r)}$. 
As the distance to the singular matrix decreases \cite{dongarra1979linpack}, this sequence of approximates approaches the true solution.
By construction we are guaranteed (in the worst case) to compute the steady state value in a maximum number of steps given by $k=\Delta_r+1$, but, in practice, we typically have $k\ll \Delta_r+1$.

Figure~\ref{fighank} highlights the contrast between the actual number of steps needed for $\epsilon$-convergence of the dynamical iteration (i.e., the dynamic simulation is within an $\epsilon$-distance of the true value, or $|x_k(r)-x^*(r)|\leq \epsilon$), and the significantly fewer number of steps needed for $\epsilon$-convergence of the Hankel approximation to the final value (i.e., $|h_k(r)-x^*(r)|\leq \epsilon$). For practical applications, these state values could be acquired from sensor measurements. For most illustrations below, we  compute these initial state values via a short simulation of the dynamical system given by \eqref{eqnconsdis_main}. We then show that the number of steps required to compute the final value of each node, in almost all cases, is significantly less than the number of steps needed for convergence of dynamical system.

We highlight several important features of this example: 
\begin{itemize}
\item Convergence of the Hankel approximation is guaranteed and upper-bounded by the Hankel rank of each node, yet in the consensus dynamics example of Figure~\ref{fighank}, the number of steps given by the upper bound was never required---and was significantly less than the number of steps needed for the corresponding dynamical iteration;
\item Each node exhibits a distinct number of steps required to compute the common consensus value---hence some nodes are in a sense more `knowledgeable' than others, and can compute or predict the final value sooner;
\item In the case of consensus models such as the one considered in Figure~\ref{fighank}, a single node, without any information other than a short sequence of its own state values, can determine the final consensus value of the \emph{entire} network.
\end{itemize}
These powerful features, and their potential wide-ranging applications, will be explored in the following section. 
\subsection*{Node-specific Predictive Capability in Complex Networks \label{nodeconv}}

\begin{figure}[t!]
\centering
\includegraphics[width=7cm]{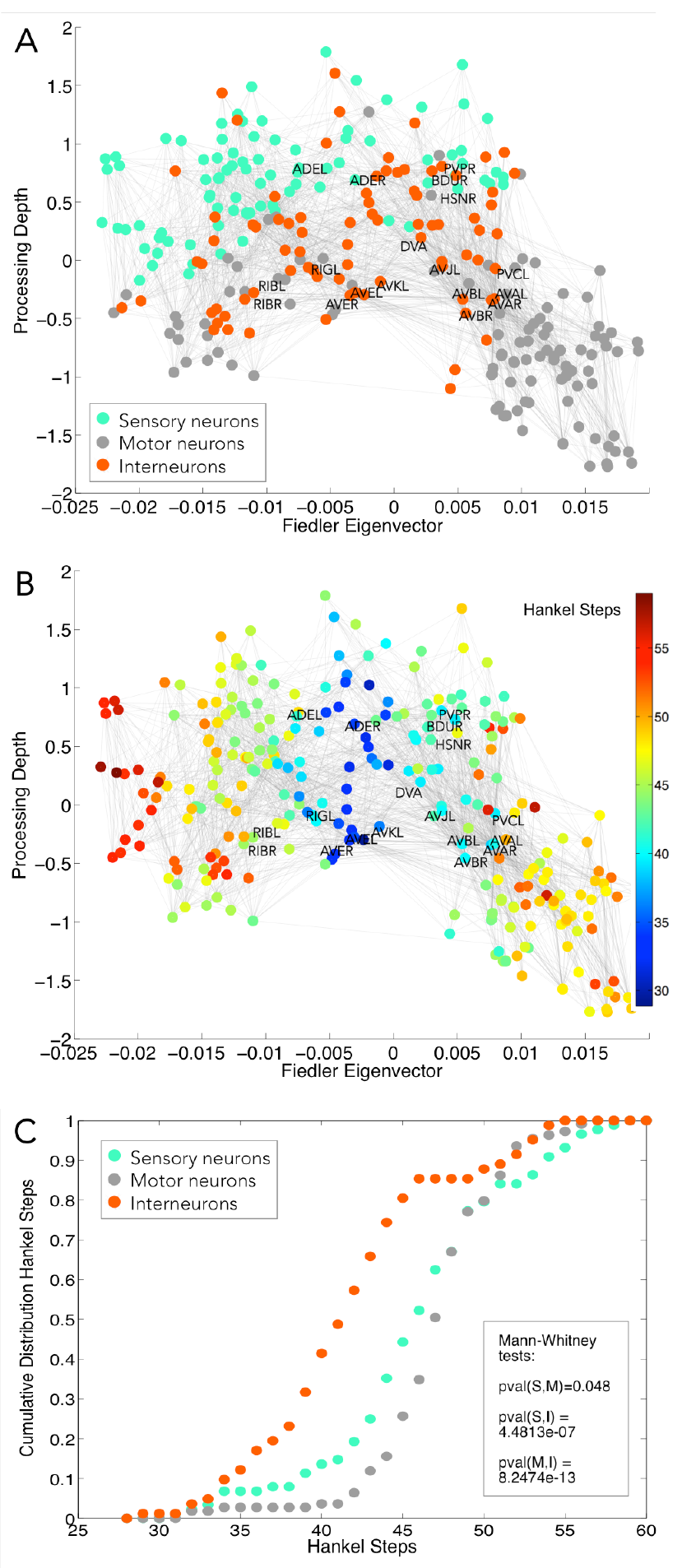}
\caption{\footnotesize \emph{Node-specific predictive capability.} 
\textbf{(A)} The well-known neuronal network of the \textit{Caenorhabditis Elegans} worm represents the synaptic connections and chemical junctions between sensory (green), motor (grey) and inter- (orange) neurons~\cite{varshney2011structural}.
The layout follows Varshney \emph{et al} \cite{varshney2011structural}: the $y$-coordinate is given by the processing depth, and the $x$-coordinate is the node coordinate of the normalised Fiedler eigenvector. 
\textbf{(B)} Each node of the network is coloured according to the number of Hankel steps it needs to approximate its own steady state value to a small tolerance ($\epsilon=10^{-4}$). The values are averaged over dynamics started from $5000$ random initial conditions in the interval $[-1,1]$. We observe that inter-neurons typically compute their steady-state value first. \textbf{(C)} There is a statistically significant difference in the number of Hankel steps needed by the three neuron groups.}  
\label{figCelegans}
\end{figure} 

The example in the previous section highlights the fact that each node of the network requires a unique number of initial state values to compute its own steady state solution, and that some nodes are better predictors in the sense that they require fewer values. Here we investigate this result by applying our approach to the well-known \emph{C. Elegans} neuronal network which describes the wiring structure of sensory, motor and inter-neurons for the \textit{Caenorhabditis Elegans} worm. \emph{C. Elegans} has been used extensively as a model organism to study neuronal development \cite{watts1998collective} as it is one of the simplest organisms with a nervous system, and is easy to grow in bulk populations. 

The \emph{C. Elegans} network is shown in Figure~\ref{figCelegans}~A, with sensory (green), motor (grey) and inter- (orange) neurons located in distinct regions of the network. Key neurons, such as interneurons responsible for mediating signals between input sensory neurons and motor neurons, are labelled explicitly on the network. The edges represent a combination of gap junction (electrical) connections and chemical synapse connections. Following \cite{varshney2011structural} nodes have been positioned with processing depth (i.e., the number of synapses from sensory to motor neurons) on the y-axis, and the node's respective entry in the Fiedler eigenvector on the x-axis. The Fiedler vector is the eigenvector corresponding to the second smallest eigenvalue of the Laplacian matrix of a graph. The entries of this vector have previously been used to partition the network into two densely connected groups of nodes \cite{newman2006finding} (positive values of the Fiedler vector in one group; negative values in the other), and for spectral embedding, i.e., for finding an optimum one-dimensional representation of a graph \cite{juvan1992optimal}.

We model the communication abilities of the neurons via a simple consensus model, where neurons (nodes) communicate with their neighbours in the network via chemical and electrical signalling \cite{varshney2011structural, macosko2009hub}. We seek to identify nodes which, after receiving a limited number of signals from their neighbours, can predict the final consensus value of all nodes or neurons in the fewest steps. These nodes, we propose, are highly knowledgeable---in the sense that they are influential communicators---about the network dynamics given their position in the network. Figure~\ref{figCelegans}~B shows the network coloured by the number of Hankel steps needed by each node to approximate its own steady state value (the consensus value of the network). These values are averaged over 5,000 random initial conditions in the interval $[-1,1]$, and detected with tolerance $\epsilon=10^{-4}$. We observe that inter-neurons, such as AVER/L, AVKL and RIGL (associated with locomotion in response to stimuli) and sensory neurons ADEL/R located in the head, typically compute their steady-state value first. Overall, inter-neurons exhibit a statistically significant decrease in number of consecutive values needed to approximate the final steady-state value, shown via comparing the three distinct neuron groups in Figure~\ref{figCelegans}~C.

Hence, key inter-neurons, having collected a small number of their own state values based on signals from their neighbours, can compute the common consensus value of the whole network, and broadcast or communicate it to other nodes who have not yet been able to compute this consensus value. In the light of this, it appears that inter-neurons have an increased predictive ability compared to more peripheral nodes, consistent with their functional role as key communicators between sensory and motor neurons in the network.

\begin{figure*}[t!]
\centering
\includegraphics[width=17cm]{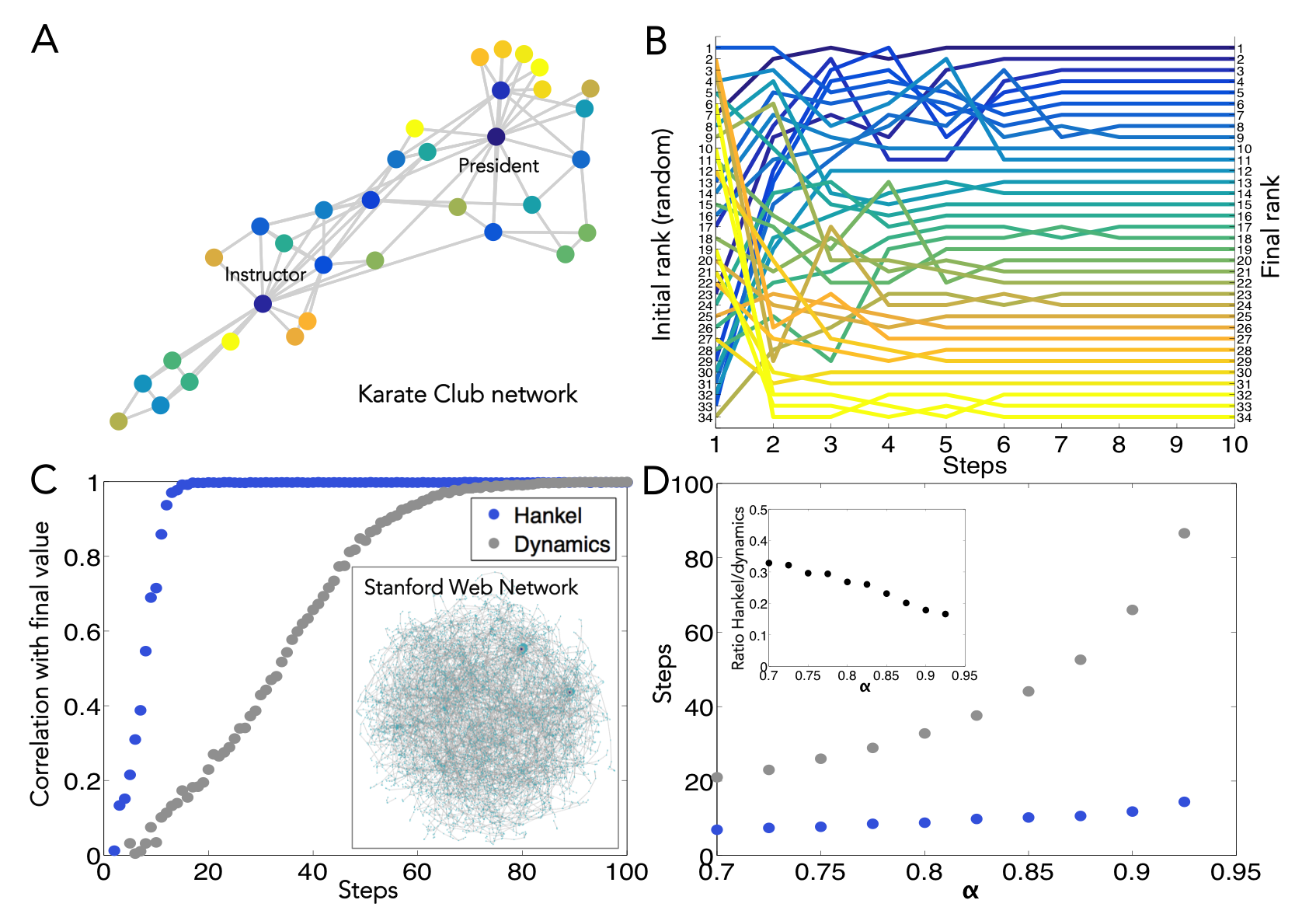}
\caption{\footnotesize \emph{Node ranking from local information.}
\textbf{(A)} The network of the \emph{Karate Club} social network \cite{karate} represents a friendship network within a US karate club dominated by two cliques---one allied to the \emph{President}, and the other to the \emph{Instructor}. The nodes are coloured by their node ranking (darker nodes are ranked higher) confirming the influence of these two key actors.
\textbf{(B)} Starting from a random initial condition (the random initial ranking on the left y-axis), a new node ranking can be obtained for each Hankel step (x-axis). By step 8, all nodes converge to their final ranking (colours of the lines as in A). 
\textbf{(C)}  The node ranking of the largest connected component of the $2002$ Stanford web network  
($n=8929$ nodes, inset)~\cite{leskovec2009community} is computed using the Hankel approximation (with damping factor $\alpha=0.9$, blue circles), and with the full iteration method (grey circles) starting from a random initial condition.
Spearman correlation of both rankings against the ground truth (i.e., the Pagerank achieved asymptotically).
The Hankel approach produces an accurate ranking (i.e., the correlation reaches $1$) in significantly fewer steps than the convergent dynamics. 
\textbf{(D)} The number of steps required by the Hankel method compared to the full iteration method when the damping factor $\alpha$ is varied between $0.7$ and $0.925$ (shown is an average over $10$ random initial conditions for each value of $\alpha$ with tolerance $\epsilon=10^{-5}$). 
For higher values of $\alpha$ the approximation is closer to the `true' ranking for an undamped system with $\alpha=1$. Hence both the iteration and the Hankel method require more steps to converge. 
(Inset) The ratio of Hankel steps to full dynamical iteration steps decreases with increasing $\alpha$, i.e., the Hankel method is increasingly more efficient as we approximate the undamped network ranking. 
\label{figdir} 
}
\end{figure*} 
\subsection*{Global Node Ranking Based on Local Information}

Beyond the consensus framework, the Hankel approximation method may be applied to a wide range of linear models. Here we consider node ranking, which forms the basis of the Google search engine technology \cite{brin1998anatomy}. Analogous to the previous example, using our approach, individual nodes can compute their own ranking while employing significantly fewer iteration steps than traditionally needed---and without the requirement of knowing the full network structure.

The classic PageRank algorithm \cite{googlebook} is based on a model of a random walker moving from node to node along the edges of a network formed by hyperlinked websites. A node's ranking can be seen as the long-run probability of the walker traversing that node relative to other nodes. The PageRank vector (i.e., the node ranking vector) is obtained as the solution of the linear system given in~\eqref{eqnconsdis_main} with $W=\alpha AD^{-1}+\frac{1-\alpha}{n} E$ where $E$ is a matrix of ones. The second term may be seen as a cost or damping term for which choices of $\alpha$ close to 1 yield the most accurate ranking---yet are more computationally expensive in the sense that convergence is slower. In essence, a value of $\alpha$ less than 1 adds constant edges to the network enabling information to diffuse and spread more quickly. For smaller values of $\alpha$, these edges are more heavily weighted and the system reaches an equilibrium state faster.

While details of the state-of-the-art Google ranking algorithms are unavailable, for large networks (the largest being the whole internet), webpage ranking employing this classic algorithm is normally approximated via $50-100$ iterations of the full system, and has been reported to take a few days to complete each month \cite{bryan200625}. Here we show that, for both a well-known social network and a large web network, we can use our Hankel method to compute the ranking value for each node in a localised and efficient manner compared to the traditional approach of iteration of the full dynamics.
\begin{figure*}[t!]
\centering
\includegraphics[width=17.5cm]{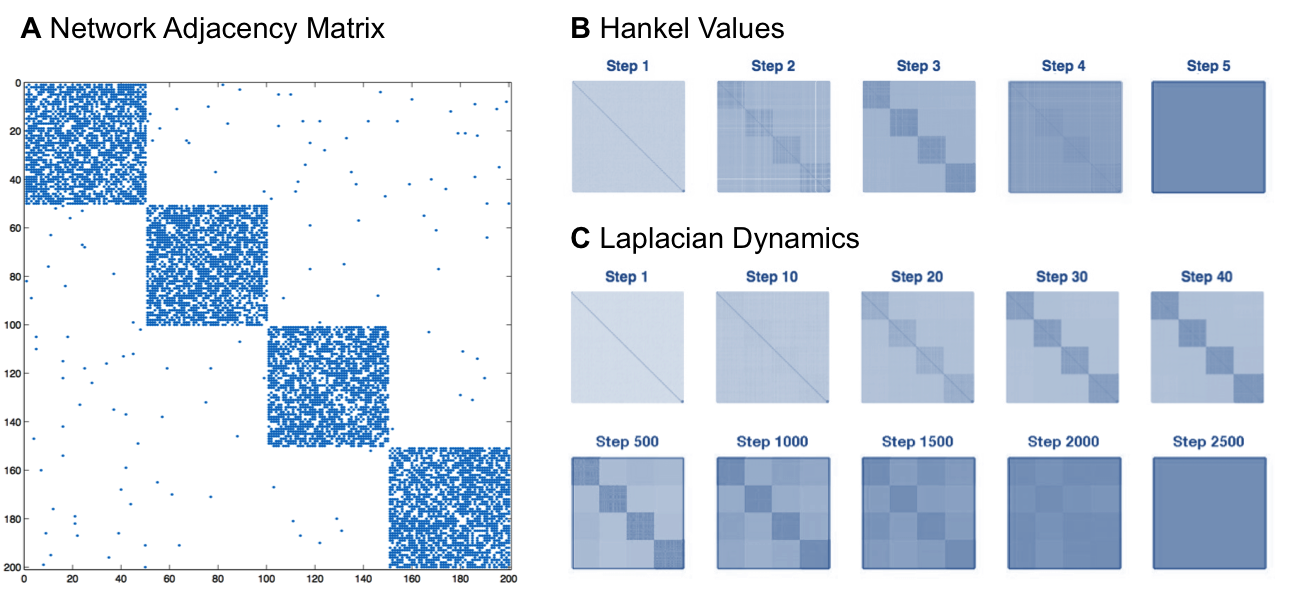}
\caption{\footnotesize 
\emph{Community Detection.}
\textbf{(A)} Adjacency matrix of a network with $n=200$ nodes and four equal sized communities (the probability of connection for node pairs within the same community is $P_{i,j}=0.7$, tand across different communities is $P_{i,j}=0.01$).
\textbf{(B)} The Hankel approximation method detects transient clustering of node dynamics, corresponding to the detection of community structure. We visualise the distance matrix $D_k$, where entries correspond to the difference between the Hankel approximations for each node pair at step $k$. The blocks of small values (dark shading) emerging at steps $2$, $3$ and $4$ correspond to the transient clustering of the Hankel approximation within communities. By step 5, the iteration has converged and all nodes have the same value ($D_k$ has only small values).
\textbf{(C)} The corresponding dynamical matrices $S_k$, where entries correspond to the distance between dynamical iteration values for node pairs at step $k$, also detects communities, but at much longer times than the Hankel approach (e.g., community structure only starts to appear at around step $20$). 
While the Hankel method computes the final consensus value in just $5$ steps, the full system dynamics takes up to $2500$ steps to converge. \label{figcomm}}
\end{figure*} 
Figure~\ref{figdir}~A illustrates the well-known Karate Club network \cite{karate}. This is a social network of friendship links between 34 members of a karate club at a US university in the 1970s. A disagreement between the club's president and main instructor created a split between the members, and ultimately led to the breakup of the club. The nodes are coloured according to the true ranking (dark blue corresponds to high ranking). In Figure~\ref{figdir}~B, for a random initial condition (corresponding to a random initial ranking shown on the left y-axis), we show the node ranking for each step of the Hankel ranking algorithm. We observe that the final ranking is achieved by most nodes by step $k=6$, and all nodes by step $k=8$. The president and instructor are the top-ranked nodes, using their dense friendship links to other club members to cement their position in rival camps. 

In order to illustrate the efficiency of our method on a larger scale, we compare the number of steps needed to obtain the node ranking of a large web network, the undirected largest connected component of the Stanford web network which describes hyperlinks between almost $10,000$ web pages in the domain stanford.edu for the year $2002$  with $n=8929$ nodes, and over $25,000$ edges~\cite{leskovec2009community}. Using a damping factor value of $\alpha=0.9$, we compute the node ranking via both the Hankel approximation method and, for comparison, iteration of the full dynamical system. In order to compare the speed of each of these approaches, we compute the Spearman rank correlation \cite{press1982numerical} between the approximate ranking (Hankel or dynamical iteration) and the true ranking (previously computed via a long iteration) at each step. Figure~\ref{figdir}~C shows that our Hankel approximation can obtain the node ranking (i.e., the correlation reaches 1) in significantly fewer steps than the linear dynamic iteration.

For large networks, the quality of this approximation depends on the choice of damping factor $\alpha$ \cite{damping}---the closer $\alpha$ is to $1$, the higher the accuracy of the approximation but also the slower the convergence rate \cite{extrapagerank} ($\alpha=0.85$ is the most commonly used value in the literature~\cite{googlebook}). Figure~\ref{figdir}~D shows that, as we increase $\alpha$ and, thereby, the accuracy of the ranking, both the iteration and the Hankel method require more steps to converge. The inset shows that the ratio of Hankel steps to dynamics steps decreases with increasing $\alpha$, implying that the Hankel method gains in relative efficiency as the damping increases. 

These results are important as many applications today contain millions, or sometimes even billions, of nodes. The potential to both locally and efficiently determine metrics such as node rankings, or equivalently a variety of centrality or dynamics-based measures, using our newly proposed method, could render previously intractable problems solvable.

\subsection*{Community Detection from Local Node Transients}

Beyond consensus models and node ranking for networks, linear systems models form the basis for a large class of computational algorithms for the analysis of network structure. Due to the large-scale, complex nature of real-world networks, the detection of communities or groups of nodes, typically tightly connected, can yield powerful insights into network behaviour by revealing the underlying organisation of the network, or providing insight into its function. Furthermore, network size may be reduced via aggregation of nodes in communities, often yielding more a tractable and informative topology \cite{yaliraki2007chemistry, Delvenne2010, oclery}.

Many algorithms have been proposed for the analysis of community structure in graphs \cite{Newman2003,f51} including normalised cut \cite{f52}, modularity \cite{f53,f54} and stability \cite{Lambiotte09laplaciandynamics,Delvenne2010}. Many of these algorithms are based on the idea that a random walker on a network (i.e., a walker that travels from node to node) becomes `trapped in wells', circulating within sub-regions of the network. This can be due, for example, to a region of high connectivity leading the walker to repeatedly traverse the same set of nodes for an extended period of time. 

The probabilistic dynamics of a random walker on a graph may be modelled via the construction of a linear system similar to that introduced above in the webpage ranking example \cite{Lambiotte09laplaciandynamics,Delvenne2010}, i.e., $W=AD^{-1}$. Transient clustering of these dynamics provides evidence of community structure in the network as the dynamics of nodes in the same community $\epsilon$-converge temporarily (i.e., they exhibit similar state values) before reaching a (possibly but not necessarily common) steady state. We find that beyond approximating the steady state solution, our Hankel approach can also detect these transient states. This important feature of our method means that, even before computing the final state for each node, Hankel approximations converge for nodes within the same community, thereby allowing communities to be very quickly identified during the Hankel iterations. Hence, as we will see, applying the Hankel method enables us to also detect community structure in a localised manner, and using significantly less successive state values than existing methods. 

To illustrate this result, we generate an ensemble of networks with community structure defined by a matrix of probabilities $P$ such that $P_{l,m}$ is the probability of connection between any node in community $l$ and any node in community $m$. Figure~\ref{figcomm}~A shows the entries in the adjacency matrix of such a network, with $n=200$ nodes split into four equal size communities (each with $50$ nodes). In this case, the probabilities of node connection between communities is given by $P_{l,m}=0.7$ for any $l=m$ (i.e., nodes in the same community) and $P_{l,m}=0.01$ for all $l\neq m$. 

For each network we compute the iterative dynamics (using $W=AD^{-1}$ and a random initial condition in the interval $[0,1]$ for \eqref{eqnconsdis_main}), and compute a sequence of Hankel values as defined by \eqref{eqnfv}. We seek to detect `distances' between the Hankel values, and the iterative dynamics, for pairs of nodes at each step. We compute distance matrices $D_k$ and $S_k$ at each step $k$ of the iteration, with entries 
\begin{align*}
D_{k}(i,j)&=|h^*_k(i)-h^*_k(j)|  \\
S_{k}(i,j)&=|x_k(i)-x_k(j)|
\end{align*}
for all pairs of nodes $i$ and $j$, where $h^*_k(i)$ is defined in \eqref{eqndecent_main} and $x_k(i)$ is the ith component of the vector ${\bf x}_k$ in \eqref{eqnconsdis_main}. Groups of nodes exhibiting similar dynamics (as captured by small distance values) signal community structure.

In Figure~\ref{figcomm}~B and C, we visualise the matrices $D_k$ and $S_k$ for increasing values of the iteration number $k$. In both cases, dark shading indicates small distance values---and hence the detection of community structure. We observe that the Hankel approximation transiently detects four communities within just $k=3$ steps, before closely approximating the final steady state value at $k=5$. In contrast, the full dynamics only detect the community structure after around $k=20$ steps, and does not converge to the consensus value until $k>2000$---illustrating the substantial advantage of the Hankel approximation for sparsely connected networks. 

\section*{Discussion}

Using local information to obtain global properties from interconnected dynamical systems. our work exhibits a number of distinguishing features. Its localised formulation, whereby nodes need only have access to a limited number of their own successive state values, enables nodes or sensors to independently compute variables of interest, without knowledge of the network structure or of their neighbouring nodes' state values. Its formulation is simple, and is easily adaptable to a large class of network-based dynamical models. It can be scaled to cope with large real-world systems, with significant efficiency compared to commonly used iterative methods as it avoids the need for convergence. Finally, it enables us to identify functional attributes of nodes in networks, i.e., `predictor' or `communicator' nodes that can estimate the full network dynamics ahead of other nodes.

Our algorithm performs well for a range of networks and applications; yet alternative but related approaches could be explored for future work. While we have employed the singular vector corresponding to the smallest singular value as an approximate nullspace vector, it may be possible to use a combination of singular vectors defining a closest subspace in order to obtain  `smoother' convergence to the consensus value. Furthermore, computing the SVD or rank update for the Hankel matrix as new rows and columns of data are added (without re-computing the decomposition) in the line of thought explored by \cite{eisenstat1995relative,brand2002incremental} would be beneficial. A closed-form solution for the updated singular vector dependent only on the previous step and the new data does not appear possible using current techniques, but could be feasible under a re-formulation or further relaxation of the problem.

For any such algorithm, it is desirable to have a fully localised convergence criterion that can be computed by each node locally to ascertain when the consensus value is well approximated. In the SI, we show that the difference between successive steps of the algorithm is highly correlated with the `distance' to the true consensus value; hence this criterion can be used as an effective stopping criterion for large graphs. Developing a theoretical convergence criterion would be beneficial, although this is a non-trivial task due to the non-monotonicity of the sequence of steady-state value approximations.

The number of steps needed for an individual agent to approximate its own final value is node-specific, and nodes that require the fewest steps can be seen as key communicators in a network. In the case of consensus dynamics, a subset of influential nodes can `predict' the final value earlier, and potentially communicate that result to other agents in the network. There is much scope to investigate further how these results could be used dynamically by nodes to aid or disrupt consensus, and to develop further applications within the context of network design and autonomous sensing.

\matmethods{

\begin{itemize}
\item The adjacency matrix, node type and node position (x and y axis co-ordinates) for the \emph{C. Elegans} neural network  \cite{varshney2011structural} in Fig.~\ref{figCelegans} may be requested from Lav Varshney (http://varshney.web.engr.illinois.edu/). 
\item The adjacency matrix for the Karate Club \cite{karate} network in Fig.~\ref{figdir} A downloaded from http://www-personal.umich.edu/~mejn/netdata/, and visualised with Gephi (https://gephi.org/).
\item The adjacency matrix for the 2002 Stanford web network \cite{leskovec2009community} in Fig.~\ref{figdir}C downloaded from https://snap.stanford.edu/data/web-Stanford.html, and visualised with Gephi (https://gephi.org/).
\end{itemize}
}

\showmatmethods 

\acknow{N.O'C. was funded by a Wellcome Trust Doctoral Studentship. G-B.S. acknowledges the support of EPSRC through the EPSRC Fellowship for Growth EP/M002187/1. M.B. acknowledges support from EPSRC Grants EP/I017267/1 and EP/N014529/1.}

\showacknow 

\bibliography{Manuscript.bbl}

\clearpage

\begin{centering}
{\Large \sf Supplementary Information} 
\end{centering}
\\
\section{Theoretical Framework}
We consider linear dynamics where ${\bf{x}}_{k} \in \mathbb{R}^n$ is a vector containing the state of each node at step $k$,  
\begin{equation}
\label{eqnconsdis}
{\bf{x}}_{k+1}=W{\bf{x}}_k
\end{equation}
for initial condition ${\bf{x}}_0$. When $W$ has a simple eigenvalue at $1$ and all the other eigenvalues within the unit disk this system is guaranteed to reach a steady state solution.

Under certain conditions the final value or steady state may be computed by each variable from a finite number of initial steps or state values \cite{Sundaram:2007fk,Sundaram2007}.
In this case, the number of steps needed is given by the polynomial degree---in particular the coefficients of the minimal polynomial for the graph. Below we briefly review the mathematics behind this approach. 
\\

The {\bf minimal polynomial} of matrix $W \in \mathbb{R}^{n\times n}$ with respect to node $r$ is the unique monic polynomial $q_r$ with minimal degree $\Delta_r + 1$
\begin{equation}
\label{eqnmin}
q_r(z) =z^{\Delta_r+1} + \sum_{i=0}^{\Delta_r} \alpha_i^r z^i  = 0
\end{equation}
such that ${\bf e_r}^Tq_r(W)={\bf 0}$, where ${\bf e_r} \in \mathbb{R}^n$ is the vector of zeros with a single 1 in position $r$.

By application of the Final Value Theorem:
$$
\label{eqnlim}
x^*(r)  = \lim_{k\rightarrow \infty}x_k(r) = \lim_{z \to 1}(z-1)X_r(z) = \frac{F_r(1)}{p_r(1)} 
$$
where $X_r(z)$, $F_r(z)$ and $p_r(z)$ are given by
\begin{eqnarray*}
X_r(z) & = & \sum_{k=0}^{\infty} x_k(r) z^{-k} \\
F_r(z) & = & \sum_{j=0}^{\Delta_r}x_j(r)z^{\Delta_r+1-j}+ \sum_{i=1}^{\Delta_r}\alpha_i^r \sum_{j=0}^{i-1}x_j(r)z^{i-j}, \text{ and }\\
p_r(z) & = &\frac{1}{z-1}q_r(z) = z^{\Delta_r}+\sum_{i=0}^{\Delta_r-1}(1+ \sum_{j=i+1}^{\Delta_r} \alpha_j^r )z^i.
\end{eqnarray*}
In this case $x^*(r)$ is given by 
\begin{equation}
\label{eqnlim2}
x^*(r)  = \frac{[x_0(r) \hdots x_{\Delta_r}(r)]{\bf w}_r}{{\bf 1}^T{\bf w}_r}
\end{equation}
where 
$$
{\bf w}_r = 
\left[ {\begin{array}{c}
1+\sum_{i=1}^{\Delta_r} \alpha_i^r \\
1+\sum_{i=2}^{\Delta_r} \alpha_i^r \\
\vdots \\
1+ \alpha_{\Delta_r}^r\\
1
\end{array} } \right] \in \mathbb{R}^{\Delta_r+1}.
$$

All that is needed in this case to compute the final value $x^*(r)$ for node $r$ is the state values for node $r$ up to step $\Delta_r$ ($\Delta_r+1$ steps in total) and the coefficients $\alpha_i^r$ for $i=1,..,\Delta_r$.
The coefficients $\alpha_i^r$ can be computed by considering the system
\begin{equation*}
\label{eqnobs}
\left[ {\begin{array}{rrrrrr}
\alpha_1^r &   \hdots &    \alpha_{\Delta_r}^r & 1
\end{array} } \right]
\left[ {\begin{array}{l}
{\bf e}_rI \\
{\bf e}_rW \\
\vdots \\
{\bf e}_rW^{\Delta_r}
\end{array} } \right]=0.
\end{equation*}
This matrix, denoted $\mathcal{Q}_r^{(\Delta_r+1)}$, is the discrete observability matrix with $\Delta_r+1$ rows \cite{Ogata:2009fk}. 
By forming the discrete observability matrix with respect to node $r$ with $k$ rows
$$
\mathcal{Q}_k^{(r)}
=
\left[ {\begin{array}{l}
{\bf e}_rI \\
{\bf e}_rW \\
\vdots \\
{\bf e}_rW^{k-1}
\end{array} } \right]
\in \mathbb{R}^{k \times n}.
$$
and increasing $k$ until the matrix loses rank at $k=\Delta_r+1$, the coefficients $\alpha_i^r$ can be obtained from the left kernel vector \cite{Sundaram:2007fk,Sundaram2007}. \footnote{Due to the Cayley-Hamilton Theorem, if $\mathcal{Q}_{\Delta_r}^{(r)}=\Delta_r$
then $\text{rank}(\mathcal{Q}_{\Delta_r+1}^{(r)})=\Delta_r$.} 

This approach however, is not decentralised as knowledge of the network structure (e.g., $W$) is required to compute the final value.
Yuan \emph{et al.} \cite{Yuan2013} developed a new approach that employs a Hankel matrix---which does not rely on $W$---to compute the $\alpha_i^r$ terms and derive a fully decentralised analogous method for the approximation of $x^*(r)$ (see \cite{Yuan2013} for more details).

\section{Hankel Method Algorithm}
\label{sect3}

In the main text we proposed an algorithm to compute a sequence of approximations to the final value of a individual variable or node by considering the singular vector corresponding to the smallest singular value of a Hankel matrix of increasing size. 

Specifically, if 
\begin{equation}
\label{hankeldef}
H_k^{(r)} =
\left[ {\begin{array}{cccc}
x_1(r)-x_0(r) & \hdots & x_k(r)-x_{k-1}(r)\\
x_2(r)-x_1(r) & & x_{k-1}(r)-x_{k-2}(r)\\
\vdots & & \vdots \\
x_k(r)-x_{k-1}(r) &   \hdots & x_{2k-1}(r)-x_{2k-2}(r) 
 \end{array} } \right], 
\end{equation}
and ${\bf v}_k^{(r)}$ is the singular vector corresponding to the smallest singular value of $H_k^{(r)}$, then the Hankel approximation for the final value of node $r$ at step $k$ is given by
\begin{equation}
\label{eqndecent}
h_{k}^*(r)=\frac{[x_0(r)\hdots x_{k-1}(r)]{\bf v_k^{(r)}}}{{\bf 1}^T{\bf v}_k^{(r)} }.
\end{equation}
The steps to compute a sequence of final value approximations for node $r$ include:
\begin{enumerate}
\item \textit{Initialisation:} Set counter $k=1$. 
\item \textit{Iteration:} For each $k$
\begin{itemize}
\item Build the Hankel matrix $H_k^{(r)}\in \mathbb{R}^{k \times k}$ given by \eqref{hankeldef}. 
\item Compute the SVD and set the singular vector corresponding to the smallest singular value equal to ${\bf v}_k^{(r)}$. 
\item Calculate the approximation $h_{k}^*(r)$ given by \eqref{eqndecent}.
\end{itemize}
\item \textit{Termination:} If $1<k\leq n$ and $|h_{k}^*(r)-h_{k-1}^*(r)|<\nu$ for some tolerance $\nu$ (see below for a discussion), then an approximate final value has been found, otherwise increment $k=k+1$ and return to step 2\footnote{For many of the examples in the main paper, where $x^*$ was known, the criterion for counting the number of Hankel steps required for the method to converge for the full network was $||h_{k}^*-x^* ||_2<\epsilon$.}. 
\end{enumerate}
By construction we are guaranteed to compute the consensus value in a maximum number of steps $k=\Delta_r+1 \leq n$.
\begin{figure}[t!]
\centering
\includegraphics[width=8.5cm]{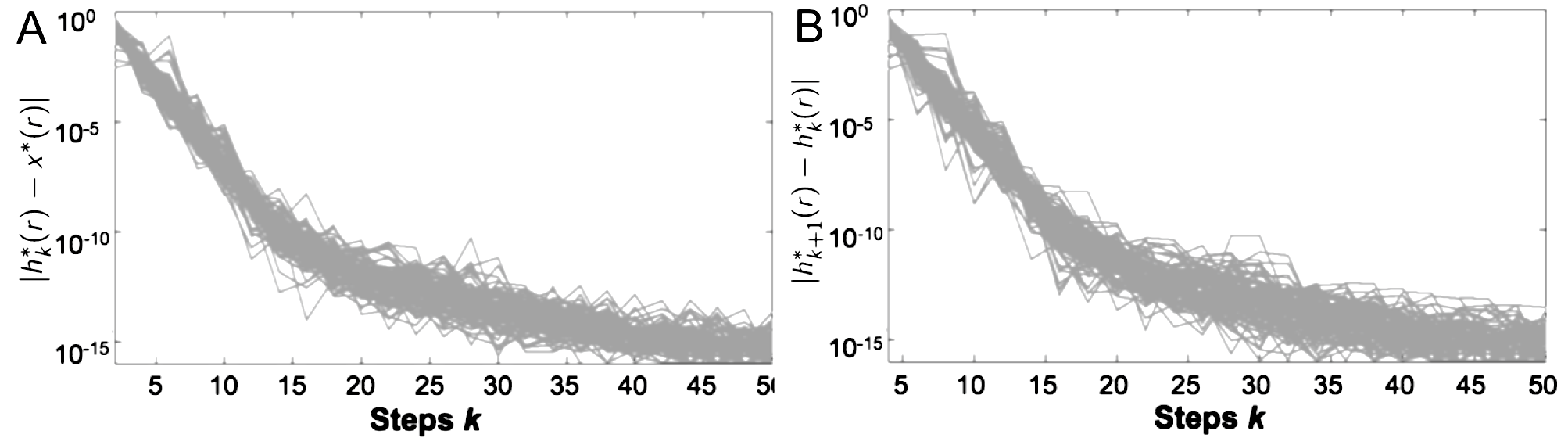}
\caption{
\textbf{(A)} For an Erd\H{o}s-R\'{e}nyi random graph of size $n=100$ and edge density $f=0.5$, we observe the convergence of the Hankel approximation for each node $r$ to the final value $x^*(r)$. 
\textbf{(B)} While the Hankel approximation converges to the final value, so do the differences between successive steps. 
}
\label{figstop}
\end{figure} 
\begin{figure*}[h!]
\centering
\includegraphics[width=18cm]{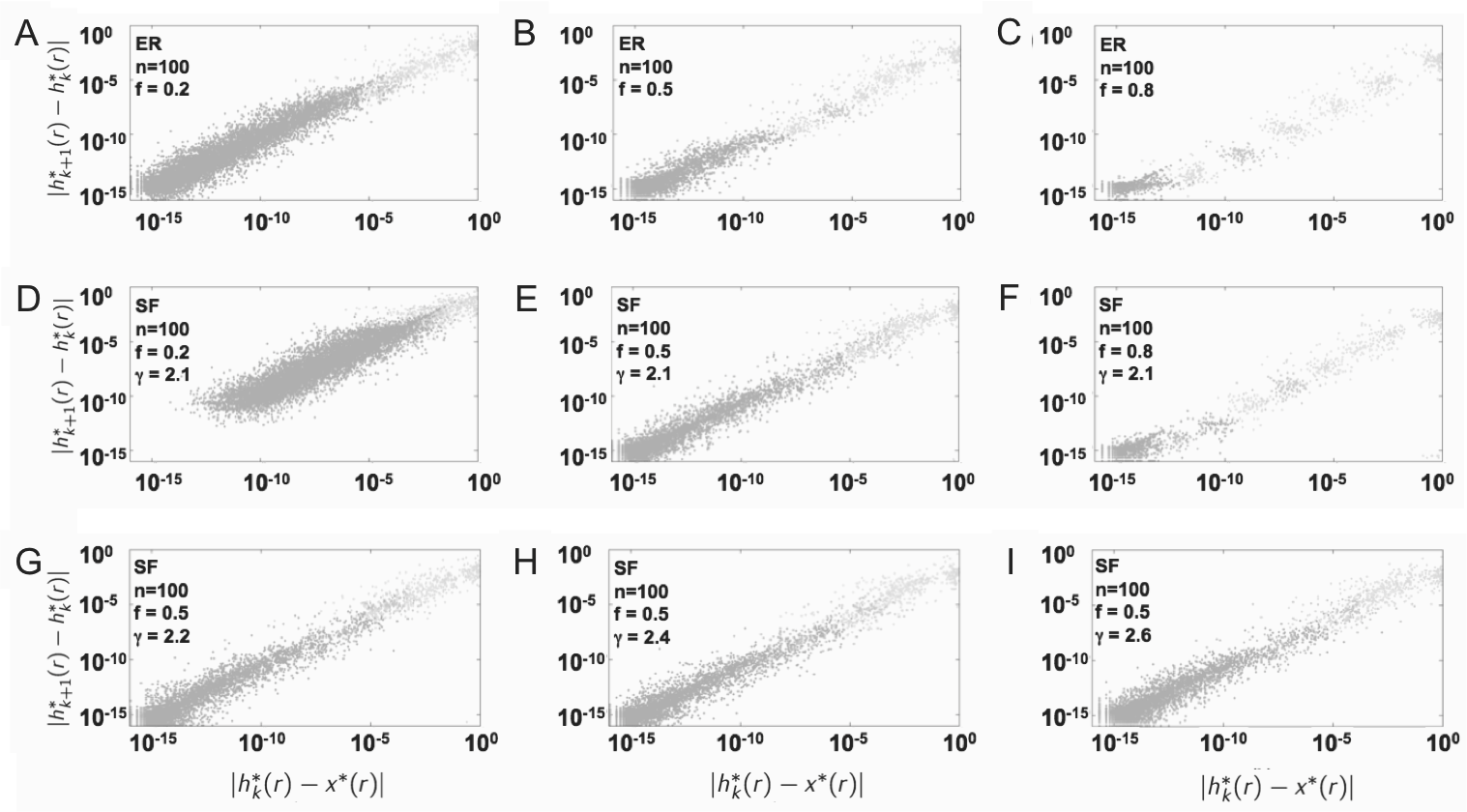}
\caption{
We explore the the relationship between the approximated vs. true consensus value, $|h^*_{k}(r)-x^{*}(r)|$, on the one hand, and the change in the approximated consensus value from one step to the next,  $|h^*_{k+1}(r)-h^*_{k}(r)|$, on the other. 
Here we observe good correlation between these quantities over variation in edge density for both Erd\H{o}s-R\'{e}nyi and scale free graphs, illustrating the appropriateness of the latter as stopping criteria for our algorithm for various kinds of large graphs, and a range of key graph parameters.
Note, the paler colour dots correspond to the first few approximation steps (corresponding to higher values of the distances on both the x and y axis).
{\bf A-C} Erd\H{o}s-R\'{e}nyi graphs of size $n=100$ with edge densities $f$ equal to $0.2, 0.5$ and $0.8$ respectively.
{\bf D-F} Scale-free graphs, which display variation in degree heterogeneity as controlled by parameter $\gamma$, of size $n=100$ (with $\gamma = 2.1$) with edge densities of $0.2, 0.5$ and $0.8$ respectively.
{\bf G-I} Scale-free graphs of constant size ($n=100$) and edge density ($f=0.5$). We vary the degree heterogeneity $\gamma = 2.2, 2.6, 3.0$ (most heterogenous to least heterogeneous).
}
\label{figstop1}
\end{figure*} 
\section{Rate of Convergence}

For an arbitrary graph, we do not know {\it a priori} the final value of any node, and hence we must define a convergence or stopping criterion for the algorithm. 
In order to assess the convergence properties of this approach, we examine the relationship between the 'true' error, $|h^*_{k}(r)-x^{*}(r)|$, and the convergence error, $|h^*_{k+1}(r)-h^*_{k}(r)|$, for a range of large graphs with differing sizes, edge densities and degree distributions as seen in Figures \ref{figstop} and \ref{figstop1}.

In Figure \ref{figstop}, we consider an Erd\H{o}s-R\'{e}nyi random graph of size n = 100 and edge density $f = 0.5$ (the fraction of edges divided by the total possible number of edges). We observe that as the Hankel approximation converges to the final value (see in subfigure A), so do the differences between successive steps (seen in subfigure B). 
Figure \ref{figstop1} directly compares the 'distance' of the Hankel approximation to the final value (x-axis), and to it's preceding value (y-axis) for a range of network structures. These include Erd\H{o}s-R\'{e}nyi graphs of varying edge density (top row), scale free graphs of various edge density and constant $\gamma$ (a parameter used to control the distribution/heterogeniety of high and low degree nodes, middle row), and scale free graphs with varying $\gamma$ (bottom row). 
We observe a consistently strong correlation between the distance of the Hankel value to the true value, and its preceding value.

We can conclude that a small convergence error is strongly indicative that the Hankel approximation has reached a similarly close approximation of the true final value. 
Hence, we deem the algorithm(s) to have converged if 
$$
|h^*_{k}(r)-h^*_{k-1}(r)|<\nu
$$ 
for some tolerance $\nu$. Figure \ref{figstop1} tells us that, in the case of an Erd\H{o}s-R\'{e}nyi or scale graph of size $n=100$, a tolerance of $\nu=$1e-5 would yield a Hankel approximation within about 1e-5 of the true solution. 
We emphasise that this criterion is fully decentralised in the sense it can be computed at each step of the algorithm, by each individual node, without any knowledge of 'global' parameters.  

\end{document}